\documentclass[12pt]{article}
\usepackage{epsfig}
\def\be{\begin{equation}}
\def\ee{\end{equation}}
\def\bea{\begin{eqnarray}}
\def\eea{\end{eqnarray}}
\usepackage{graphicx}% Include figure files

\catcode`\@=11
\def\lsim{\mathrel{\mathpalette\@versim<}}
\def\gsim{\mathrel{\mathpalette\@versim>}}
\def\@versim#1#2{\vcenter{\offinterlineskip
\ialign{$\m@th#1\hfil##\hfil$\crcr#2\crcr\sim\crcr } }}
\catcode`\@=12
\usepackage{axodraw}

\catcode`@=12
\begin{document}
\thispagestyle{empty}
\begin{flushright}
UCRHEP-T525\\
UMISS-HEP-2013-03\\
February 2013\
\end{flushright}
\vspace{0.6in}
\begin{center}
{\LARGE \bf Radiative Scaling Neutrino Mass\\ 
with $A_4$ Symmetry\\}
\vspace{0.6in}
{\bf Subhaditya Bhattacharya$^1$, Ernest Ma$^1$, Alexander Natale$^1$,\\ and 
Ahmed Rashed$^{2,3}$\\}
\vspace{0.2in}
{\sl $^1$ Department of Physics and Astronomy, University of
California,\\
Riverside, California 92521, USA\\}
\vspace{0.1in}
{\sl $^2$  Department of Physics and Astronomy, 
University of Mississippi,\\ Oxford, Mississippi 38677, USA\\}
\vspace{0.1in}
{\sl $^3$  Department  of Physics, Faculty of Science, Ain Shams University,\\ 
  Cairo, 11566, Egypt\\}
\end{center}
\vspace{0.6in}
\begin{abstract}\
A new idea for neutrino mass was proposed recently, where its smallness is not 
due to the seesaw mechanism, i.e. not inversely proportional to some large 
mass scale. It comes from a one-loop mechanism with dark matter in the loop 
consisting of singlet Majorana fermions $N_i$ with masses of order 10 keV and 
neutrino masses are scaled down from them by factors of about $10^{-5}$. 
We discuss how this model may be implemented with the non-Abelian discrete 
symmetry $A_4$ for neutrino mixing, and consider the phenomenology of $N_i$ 
as well as the extra scalar doublet $(\eta^+,\eta^0)$.
\end{abstract}

\newpage
\baselineskip 24pt

The origin of neutrino mass is the topic of many theoretical discussions. 
The consensus is that its smallness is due to some mass scale larger than 
the electroweak breaking scale of about 100 GeV.  If there are no particles 
beyond those of the standard model (SM) lighter than this scale, then the 
well-known unique dimension-five opeartor~\cite{w79}
\begin{equation}
{\cal L}_5 = {-f_{ij} \over 2 \Lambda} (\nu_i \phi^0 - l_i \phi^+) 
(\nu_j \phi^0 - l_j \phi^+) + H.c.
\end{equation}
induces Majorana neutrino masses as the Higgs scalar $\phi^0$ acquires 
a nonzero vacuum expectation value $\langle \phi^0 \rangle = v$, so that
\begin{equation}
({\cal M}_\nu)_{ij} = {f_{ij} v^2 \over \Lambda}. 
\end{equation}
This shows that neutrino mass is seesaw in character, i.e. it is 
inversely proportional to some large scale $\Lambda$.  The ultraviolet 
completion of this effective operator may be accomplished in three ways 
at tree level~\cite{m98} using (I) heavy Majorana fermion singlets $N_i$, 
(II) a heavy scalar triplet $(\xi^{++},\xi^+,\xi^0)$, or (III) heavy 
Majorana fermion triplets $(\Sigma^+,\Sigma^0,\Sigma^0)_i$, commonly 
referred to as Type I, Type II, or Type III seesaw.  There are also 
three one-particle-irreducible (1PI) one-loop realizations~\cite{m98}. 
Recently the one-particle-reducible (1PR) diagrams have also been 
considered~\cite{bhOw12}.

If there are new particles with masses below the elctroweak scale, such as 
fermion singlets $\nu_S$ with mass $m_S$, then neutrinos may acquire mass 
through their mixing with $\nu_S$.  However, this mechanism is still seesaw 
because $m_\nu$ is still inversely proportional to $m_S$.  There is however 
an exception.  It has been pointed out recently~\cite{m12} that in the 
scotogenic model of radiative neutrino mass~\cite{m06}, it is possible 
to have $m_\nu$ directly proportional to $m_S$, and there is no mixing 
between $m_\nu$ and $m_S$.

This model was proposed~\cite{m06} in 2006 to connect neutrino mass 
with dark matter.   The idea is very simple.  Assume three neutral fermion 
singlets $N_i$ as in the usual Type I seesaw~\cite{seesaw}, but let them 
be odd under a new $Z_2$ symmetry, so that there is no $(\nu_i \phi^0 - 
l_i \phi^+) N_j$ coupling and the effective operator of Eq.~(1) is 
not realized.  At this stage, $N_i$ may have Majorana masses $M_i$, but 
$\nu_i$ is massless.  However, they can be linked through the 
interaction $h_{ij} (\nu_i \eta^0 - l_i \eta^+) N_j$ where $(\eta^+,\eta^0)$ 
is a new scalar doublet which is also odd under the aforementioned 
$Z_2$~\cite{dm78}.  Hence Majorana neutrino masses are generated in one loop 
as shown in Fig.~1.
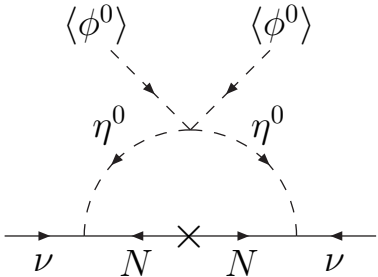
\begin{figure}[htb]
\begin{center}
\begin{picture}(260,120)(0,0)
\ArrowLine(60,10)(90,10)
\ArrowLine(130,10)(90,10)
\ArrowLine(130,10)(170,10)
\ArrowLine(200,10)(170,10)
\DashArrowArc(130,10)(40,90,180)5
\DashArrowArcn(130,10)(40,90,0)5
\DashArrowLine(100,80)(130,50)5
\DashArrowLine(160,80)(130,50)5
\Text(75,0)[]{\large $\nu$}
\Text(185,0)[]{\large $\nu$}
\Text(110,0)[]{\large $N$}
\Text(150,0)[]{\large $N$}
\Text(100,52)[]{\large $\eta^0$}
\Text(160,52)[]{\large $\eta^0$}
\Text(95,90)[]{\large $\langle \phi^0 \rangle$}
\Text(165,90)[]{\large $\langle \phi^0 \rangle$}
\Text(130,10)[]{\Large $\times$}
\end{picture}
\end{center}
\caption{One-loop generation of scotogenic Majorana neutrino mass.}
\end{figure}
This mechanism has been called ``scotogenic'', from the Greek ``scotos'' 
meaning darkness.  Because of the allowed $(\lambda_5/2) (\Phi^\dagger \eta)^2 
+ H.c.$ interaction, $\eta^0 = (\eta_R + i \eta_I)/\sqrt{2}$ is split so 
that $m_R \neq m_I$.  The diagram of Fig.~1 can be computed exactly~\cite{m06}, 
i.e.
\begin{equation}
({\cal M}_\nu)_{ij} = \sum_k {h_{ik} h_{jk} M_k \over 16 \pi^2} 
\left[ {m_R^2 \over m_R^2 - M_k^2} \ln {m_R^2 \over M_k^2} - 
{m_I^2 \over m_I^2 - M_k^2} \ln {m_I^2 \over M_k^2} \right].
\end{equation} 
A good dark-matter candidate is $\eta_R$ as first pointed out in 
Ref.~\cite{m06}.  It was subsequently proposed by itself in Ref.~\cite{bhr06} 
and studied in detail in Ref.~\cite{lnot07}.  The $\eta$ doublet has become 
known as the ``inert'' Higgs doublet, but it does have gauge and scalar 
interactions even if it is the sole addition to the standard model. 

The usual assumption for neutrino mass in Eq.~(1) is
\begin{equation}
m_I^2 - m_R^2 << m_I^2 + m_R^2 << M_k^2,
\end{equation}
in which case
\begin{equation}
({\cal M}_\nu)_{ij} = {\lambda_5 v^2 \over 8 \pi^2} \sum_k 
{h_{ik} h_{jk} \over M_k} \left[ \ln {M_k^2 \over m_0^2} - 1 \right],
\end{equation}
where $m_0^2 = (m_I^2 + m_R^2)/2$ and $m_R^2 - m_I^2 = 2 \lambda_5 v^2$ 
($v = \langle \phi^0 \rangle$).  This scenario is often referred to as 
the radiative seesaw.  What was not realized in most applications of 
this model since 2006 is that there is another very interesting scenario, 
i.e. 
\begin{equation}
M_k^2 << m_R^2, ~m_I^2.
\end{equation}
Neutrino masses are then given by~\cite{m12}
\begin{equation}
({\cal M}_\nu)_{ij} = {\ln (m_R^2/m_I^2) \over 16 \pi^2} \sum_k h_{ik} h_{jk} M_k.
\end{equation}
This simple expression is actually very extraordinary, because neutrino 
mass is now not inversely proportional to some large scale.  In that case, 
how do we understand the smallness of $m_\nu$?  The answer is lepton 
number.  In this model, $(\nu,l)_i$ have lepton number $L=1$ and $N_k$ have 
$L=-1$, and $L$ is conserved in all interactions except for the Majorana 
mass terms $M_k$ which break $L$ to $(-1)^L$.  We may thus argue 
that $M_k$ should be small compared to all other mass terms which conserve 
$L$, the smallest of which is the electron mass, $m_e = 0.511$ MeV.  It 
is thus reasonable to have $M_k \sim 10$ keV, in which case $m_\nu \sim 0.1$ 
eV is obtained if $h^2 \sim 10^{-3}$ in Eq.~(7).  Each neutrino mass is then 
simply proportional to a linear combination of $M_k$ according to Eq.~(7).  
Their ratio is just a scale factor and small neutrino masses are due to this 
``scaling'' mechanism.  Note that the interesting special case where only 
$M_1$ is small has been considered previously~\cite{as09,gop10}.  Note 
also that if $|m_I^2-m_R^2| = 2 |\lambda_5| v^2 << |m_I^2+m_R^2|$, 
then $\ln (m_R^2/m_I^2)$ would be strongly suppressed, but this is not 
compulsory.  For example, let $m_R = 240$ GeV, $m_I = 150$ GeV, then 
$|\lambda_5| = 0.58$ and $\ln (m_R^2/m_I^2) = 0.94$.

The scotogenic model~\cite{m06} with large $M_k$, i.e. Eq.~(5), has been 
extended recently~\cite{mnr12} to include the well-known non-Abelian discrete 
symmetry $A_4$~\cite{mr01,m02,bmv03}.  Here we consider the case of Eq.~(7). 
This assumption changes the phenomenology of $N_k$ as well as $(\eta^+,\eta^0)$ 
and may render this model to be more easily verifiable at the Large Hadron 
Collider (LHC).  We let $(\eta^+,\eta^0)$ be a singlet under $A_4$ and 
both $(\nu_i,l_i)$ and $N_k$ to be triplets.  In that case,
\begin{equation}
h_{ik} = h \delta_{ik},
\end{equation}\
and
\begin{equation}
{\cal M}_\nu = \zeta {\cal M}_N,
\end{equation}
where $\zeta = h^2 \ln (m_R^2/m_I^2)/ 16 \pi^2$ is the scale factor.  The soft 
breaking of $A_4$ which shapes ${\cal M}_N$ is then directly transmitted to 
${\cal M}_\nu$.

One immediate consequence of this restricted scaling mechanism for neutrino 
mass is that if $M_{1,2,3}$ are all of order 10 keV, then the three neutrino 
masses are all of order 0.1 eV, i.e. a quasidegenerate scenario. 
For example, if $m_1 = 0.1$ eV and is the lightest, then for $M_1 = 10$ keV, 
$M_3 = 10^{5} (m_1 + \sqrt{\Delta m^2_{31}}) = 14.85$ keV.  Another immediate 
consequence is that the interactions of $N_{1,2,3}$ with the charged leptons 
through $\eta^+$ depend only on $h$ and the mismatch between the 
charged-lepton mass matrix and the neutrino mass matrix, i.e. the 
experimentally determined neutrino mixing matrix $U_{l \nu}$.  Hence 
$\mu \to e \gamma$ is highly suppressed because the leading term of its 
amplitude is proportional to $\sum_k h_{\mu k} h^*_{e k} = |h|^2 \sum_k 
U_{\mu k} U^*_{e k} = 0$.  The next term $\sum_k U_{\mu k} U^*_{e k} M_k^2/m^2_{\eta^+}$ 
is nonzero but is negligibly small.  This means that there is no useful 
bound on the $\eta^+$ mass from $\mu \to e \gamma$.  {\it Note that $A_4$ may 
be replaced by any other flavor symmetry as long as it is possible to have 
Eq.~(8) using the singlet and triplet representations of that symmetry.}
As for the muon anomalous magnetic moment, it is given by~\cite{mr01-1}
\begin{equation}
\Delta a_\mu = - {m_\mu^2 |h|^2 \over 96 \pi^2 m^2_{\eta^+}} = -1.18 \times 10^{-12} 
\left( {|h|^2 \over 10^{-3}} \right) \left( {100~{\rm GeV} \over m_{\eta^+}} 
\right)^2.
\end{equation}
Since the experimental uncertaintly is $6 \times 10^{-10}$, this also does 
not give any useful bound on the $\eta^+$ mass.

If $\eta^\pm, \eta_R, \eta_I$ are of order $10^2$ GeV, the interactions of 
$N_k$ with the neutrinos and charged leptons are weaker than the usual 
weak interaction, hence $N_k$ may be considered ``sterile'' 
and become excellent warm dark-matter candidates~\cite{dvs11,dvfs12}.
However, unlike the usual sterile neutrinos~\cite{white12} which mix with 
the active neutrinos, the lightest $N_k$ here is absolutely stable.  This 
removes one of the most stringent astrophysical constraints on warm dark 
matter, i.e. the absence of galactic X-ray emission from its decay, which 
would put an upper bound of perhaps 2.2 keV on its mass~\cite{wlp12}, whereas 
Lyman-$\alpha$ forest observations (which still apply in this case) impose 
a lower bound of perhaps 5.6 keV~\cite{vbbhrs08}.  Such a stable sterile 
neutrino (called a ``scotino'') is also possible in an unusual left-right 
extension~\cite{m12-1} of the standard model.  Conventional 
left-right models where the $SU(2)_R$ neutrinos mix with the $SU(2)_L$ 
neutrinos have also been studied~\cite{bhl10,bkl12,nsz12}.

Since $N_k$ are assumed light, muon decay proceeds at tree level 
through $\eta^+$ exchange, i.e. $\mu \to N_\mu e \bar{N}_e$.  The 
inclusive rate is easily calculated to be
\begin{equation}
\Gamma (\mu \to N_\mu e \bar{N}_e) = { |h|^4 m^5_{\mu} \over 6144 \pi^3 
m^4_{\eta^+}}.
\end{equation}
Since $N_{\mu}$ and $\bar{N}_e$ are invisible just as $\nu_\mu$ and 
$\bar{\nu}_e$ are invisible in the dominant decay $\mu \to \nu_\mu e 
\bar{\nu}_e$ (with rate $G_F^2 m_\mu^5/192 \pi^3$), this would change 
the experimental value of $G_F$.  Using the experimental uncertainty 
of $10^{-5}$ in the determination of $G_F$, we find 
\begin{equation}
m_{\eta^+} > 70~{\rm GeV}
\end{equation}
for $|h|^2 = 10^{-3}$.  This is a useful bound on the $\eta^+$ mass, 
but it is also small enough so that $\eta^+$ may be observable at the LHC.
The phenomenological bound on $m_{\eta^+}$ from $e^+ e^-$ production at LEPII 
has been estimated~\cite{pt07} to be 70 -- 90 GeV.  A bound of 80 GeV 
was used in a previous study~\cite{dmst10} of this model.

Whereas the lightest scotino, say $N_1$, is absolutely stable, 
$N_{2,3}$ will decay into $N_1$ through $\eta_R$ and $\eta_I$.  The decay rate 
of $N_3 \to N_1 \bar{\nu}_1 \nu_3$ is given by
\begin{eqnarray}
&& \Gamma (N_3 \to N_1 \bar{\nu}_1 \nu_3) = {|h|^4 \over 
256 \pi^3 M_3} \left( {1 \over m_R^2} + {1 \over m_I^2} \right)^2 \nonumber \\ 
&& \times \left( {M_3^6 \over 96} - {M_1^2 M_3^4 \over 12} + {M_1^6 \over 12} 
- {M_1^8 \over 96 M_3^2} + {M_1^4 M_3^2 \over 8} \ln {M_3^2 \over M_1^2} 
\right).
\end{eqnarray}
Let $M_1 = 10$ keV, $M_3 = 14.85$ keV, $|h|^2 = 10^{-3}$, $m_R = 240$ GeV, 
$m_I = 150$ GeV, then this rate is $1.0 \times 10^{-46}$ GeV, corresponding 
to a lifetime of $2.1 \times 10^{14}$ y, which is much longer than the age 
of the Universe of $13.75 \pm 0.11 \times 10^9$ y.  The lifetime of $N_2$ 
is even longer because $\Delta m^2_{21} << \Delta m^2_{31}$.  Hence both $N_2$ 
and $N_3$ are stable enough to be components of warm dark matter.  However, 
$N_{2,3} \to N_1 \gamma$ are negligible for the same reason that 
$\mu \to e \gamma$ is negligible, so they again have no galactic X-ray  
signatures.\\

\begin{table}[htb]
\centering
\begin{tabular}{|c|c|c|c|c|c|}
 \hline 
$m_{\eta^{\pm}}$ (GeV) & $ \not\!\!E_T > 0$ GeV & $ \not\!\!E_T>25$ GeV & 
$\not\!\!E_T>50$ GeV & $ \not\!\!E_T>100$ GeV \\	\hline 
80 & 33.2 & 27.9 & 18.3   & 2.88 \\	\hline 
90 & 22.7 & 19.8 & 14.4   & 3.10 \\	\hline 
100 & 15.7 & 14.0 & 10.6  & 3.08 \\	\hline 
110 & 11.4 & 10.3 & 8.13  & 3.03 \\	\hline 
120 & 8.72 & 7.99 & 6.54  & 2.91 \\	\hline 
130 & 6.45 & 5.98 & 5.05  & 2.57 \\	\hline 
140 & 4.97 & 4.64 & 3.96  & 2.21 \\	\hline 
150 & 3.84 & 3.62 & 3.16  & 1.89 \\ \hline
 \hline 
SM Background & 626.45 & 453.3 &  205.8 &  8.6\\
\hline
\end{tabular}
\caption{Cross sections (fb) for signal and dominant SM background at LHC 
with $E_{CM}$= 8 TeV. For the signal, $pp \rightarrow \eta^{\pm} \eta^{\mp} 
\rightarrow e^{\pm} \mu^{\mp} N_1 N_2$ (fb) with various $ \not\!\!E_T$ cuts 
and different $m_{\eta}$ (GeV) are specified.  For SM background, $W^+W^-$ 
production and decays to $e^{\pm} \mu^{\mp} +  \not\!\!E_T$ with the same 
cuts are specified.}
\end{table}

Since $\eta^+$ may be as light as 70 GeV, it may be observable at the LHC. 
Assuming that the recently observed particle~\cite{atlas12,cms12} at the 
LHC is the Higgs boson $H$ coming from $(\phi^+,\phi^0)$, the decay 
$H \to \eta^+ \eta^-$ is not allowed for $m_H = 126$ GeV.  However, $\eta^\pm$ 
will contribute to the $H \to \gamma \gamma$ rate, as already pointed 
out~\cite{p11,abg12,ccty12,sk12}.   What sets our model apart is the 
inclusive decay of $\eta^\pm \to l^\pm N_{1,2,3}$, which is of universal 
strength.  At the LHC, 
the pair production of $\eta^+ \eta^-$ will then lead to $l_i^+ l_j^-$ final 
states with equal probability for each flavor combination.  For example, 
$e^+ \mu^-$ and $\mu^+ e^-$ will each occur 1/9 of the time.  This signature 
together with the large missing energy of $N_{1,2,3}$ may allow it to be 
observed at the LHC.  However, these events also come from $W^+ W^-$ 
production and their subsequent leptonic decays.  As shown in Table 1, 
the cross sections for the signal events are smaller than the dominant 
$W^+W^-$ background even after a large $ \not\!\!E_T$ cut.  This is because 
both the signal and background events have similar missing energy 
distributions, but the $W^+W^-$ production is much larger.  If data 
show an excess of such events~\cite{cmsww} over the SM prediction, it could 
be due to $\eta^+ \eta^-$, but it may also simply come from an incorrect 
scale factor used in the SM calculation.  Hence it is difficult to draw any  
conclusion for the case at hand.  We generated the signal events using 
{\tt Calchep}~\cite{calchep} and interfacing them to the event generator 
{\tt Pythia}~\cite{pythia}.  Background events were generated with 
{\tt Pythia} and an appropriate $K-$factor was applied to match the NLO 
cross-section for $W^+W^-$ at 8 TeV for the LHC, which is 57.3 pb~\cite{cmsww}. 
We used CTEQ6L~\cite{CTEQ} parton distribution functions for both signal 
and background events.  The basic $p_T >10$ GeV and $|\eta|<2.5$ cuts are 
applied for leptons.

In the supersymmetric $SU(5)$ completion~\cite{m08} of this model, there 
are exotic quarks which may be produced abundantly.  Their decays into 
$\eta^\pm$ would have four leptons of different flavor in the final 
state.  This may be a better signature of this model.  Details will be 
given elsewhere.

In conclusion, the scotogenic model~\cite{m06} of neutrino mass with  
a solution~\cite{m12} where there is no seesaw mechanism and $N_{1,2,3}$ 
have masses of order 10 keV has been implemented with the non-Abelian 
discrete symmetry $A_4$.  The scotinos $N_{1,2,3}$ are good warm dark-matter 
candidates which can explain the structure of the Universe at all 
scales~\cite{dvs11,dvfs12}.  Since $N_1$ is absolutely stable and the 
decays $N_{2,3} \to N_1 \gamma$ are negligible, the galactic 
X-ray upper bound of perhaps 2.2 keV on its mass~\cite{wlp12} is avoided.  
It will also not be detected in terrestrial experiments.  On the other hand, 
since this model requires an extra scalar doublet, and $\eta^\pm$ may be 
as light as 70 GeV, it may be tested at the LHC, especially if it is 
the decay product of an exotic quark.

\noindent \underline{Acknowledgment}~:~ This work is supported in part 
by the U.~S.~Department of Energy under Grant No.~DE-AC02-06CH11357.
The work of A. Rashed is supported in part by the National Science 
Foundation under Grant No. NSF PHY-1068052 and in part by the Graduate 
Student Council Research Grant Award 2012-2013.

%\newpage
\bibliographystyle{unsrt}

\end{document}